\documentclass[aps,prl,showpacs,english,10pt,a4,nofootinbib,notitlepage,twocolumn,superscriptaddress]{revtex4-1}
  
\usepackage{graphicx}
\usepackage{hyperref}
\usepackage{color}
\usepackage{amsthm,amsmath,amssymb,dsfont}
\usepackage{pgfplots}

\newcommand{\unit}{\mathbb{I}}
\newcommand{\tr}{\mathrm{Tr}}
\newcommand{\ct}{^{\dagger}}
\newcommand{\av}[1]{\lvert #1\rvert}
\newcommand{\norm}[1]{\lVert #1\rVert}
\newcommand{\asq}[1]{\left[ #1 \right]}
\newlength\figureheight
\newlength\figurewidth

\begin{document}

\title{Robust Characterization of Leakage Errors}
\author{Joel J. Wallman}
\author{Marie Barnhill}
\affiliation{Institute for Quantum Computing, University of Waterloo, 
Waterloo, Ontario N2L 3G1, Canada}
\affiliation{Department of Applied Mathematics, University of Waterloo, 
Waterloo, Ontario N2L 3G1, Canada}
\author{Joseph Emerson}
\affiliation{Institute for Quantum Computing, University of Waterloo, 
Waterloo, Ontario N2L 3G1, Canada}
\affiliation{Department of Applied Mathematics, University of Waterloo, 
Waterloo, Ontario N2L 3G1, Canada}
\affiliation{Canadian Institute for Advanced Research, Toronto, Ontario M5G 
1Z8, Canada}
\date{\today}

\begin{abstract}
Leakage errors arise when the quantum state leaks out of some subspace of 
interest, for example, the two-level subspace of a multi-level system defining 
a computational `qubit' or the logical code space defined by some quantum 
error-correcting code or decoherence-free subspace. Leakage errors pose a 
distinct challenge to quantum control relative to the more well-studied 
decoherence errors and can be a limiting factor to achieving fault-tolerant 
quantum computation. Here we present scalable and robust randomized 
benchmarking protocols for quickly estimating the rates of both coherent and 
incoherent leakage due to an arbitrary Markovian noise process, allowing for 
practical minimization of the leakage rate by varying over control methods. We 
illustrate the reliability of the protocol through numerical simulations.
\end{abstract}


\maketitle

In order to build a practical and universal quantum computer, the rate of 
decoherence and other errors must be below certain fault-tolerant 
thresholds.  One way of determining these error rates is to completely 
characterize the noise on quantum gates using quantum process tomography 
(QPT)~\cite{Chuang1997,Poyatos1997}. However, QPT scales exponentially in the 
number of qubits and is sensitive to state preparation and measurement (SPAM) 
errors, which can be on the same order as (or even orders of magnitude greater 
than) the error on the gate operations of interest~\cite{Merkel2013}.

An alternative to QPT is a characterization toolkit called randomized 
benchmarking (RB)~\cite{Emerson2005,Knill2008,Emerson2007,Lopez2010,
Magesan2011,Magesan2012,Gaebler2012,Magesan2012a,Kimmel2014,Wallman2014}. These 
RB protocols scale favorably with the number of qubits at the cost of obtaining 
only partial information about the decoherence and control errors. Many of 
these protocols offer the additional advantage of being insensitive to SPAM 
errors by applying random sequences of quantum operations drawn from a group 
and extracting average error parameters from the observed fidelity decay 
curves. These protocols have consequently become an important tool in the 
validation and verification of quantum operations and have 
provided an efficient method to optimize over experimental implementations~\cite{Brown2011,Barends2014,Corcoles2013}.

An important error mechanism in many experimental implementation is leakage
outside of the Hilbert space under consideration. Such leakage errors can be a 
substantial obstacle to fault-tolerant 
computation~\cite{Aliferis2007,Fowler2013,Ghosh2013,Whiteside2014}. For 
example, the surface code may not be used directly if there is any probability 
of losing a qubit, while for the topological cluster states, leakage rates of 
less than $1\%$ are required to avoid impractical 
overheads~\cite{Whiteside2014}.

However, standard RB only provides limited information about leakage rates~\cite{Epstein2014}. 
There are platform-dependent methods for characterizing leakage in 
many of the leading experimental approaches to quantum computation, such as 
ion trap qubits~\cite{JDevitt2007}, superconducting 
qubits~\cite{Zhou2005,Motzoi2009} and quantum dots~\cite{Medford2013}. 
However, these approaches do not have all the advantages of RB, in particular, 
scalability with the number of qubits, robustness to SPAM and no assumptions 
about the underlying error process beyond the assumption of Markovianity.

Furthermore, there are two distinct types of leakage, which we refer to as 
incoherent and coherent. In many physical implementations, incoherent leakage 
can be broadly categorized as the probabilistic but permanent loss of the 
system (through some process such as photon absorption, etc.); while coherent 
leakage can be considered as a coherent transition to an extra dimension 
(e.g., an electron excitation to an energy level outside the Hilbert space 
being considered) which later transitions back to the Hilbert space under 
consideration. These transitions back to the Hilbert space make coherent 
leakage a fundamentally non-Markovian process.

We present a protocol that provides an estimate of the average leakage rate 
for both coherent and incoherent leakage over a given set of quantum gates. 
We consider computational and leakage spaces of arbitrary dimensions, so that 
our protocol can be applied to both physical and logical qudit systems. We 
demonstrate that our protocol produces reliable estimates of leakage rates 
through numerical simulations of our protocol for specific error models.

\textit{Defining survival and leakage rates}---Within the broad framework of
time-dependent Markovian noise, any experimental implementation of a unitary $g$
at a time step $t$ can be written as $g\circ\mathcal{E}_{g,t}$ for some 
completely positive (CP) map 
$\mathcal{E}_{g,t}:\mathcal{B}(\mathcal{H})\to\mathcal{B}(\mathcal{H})$ where 
$\circ$ denotes composition (i.e., $\mathcal{A}\circ\mathcal{B}$ means apply 
$\mathcal{B}$ then apply $\mathcal{A}$) and $\mathcal{B}(\mathcal{H})$ is the 
set of density operators acting on $\mathcal{H}$ and $\mathcal{H}$ is the 
relevant physical Hilbert space (that is, $\mathcal{E}_{g,t}$ sends quantum 
states to quantum states). Note that if, as is often the case, $g$ acts on a 
subspace $\mathcal{H}_1$ where $\mathcal{H} = \mathcal{H}_1\oplus 
\mathcal{H}_2$, then we implicitly extend $g$ to $g\oplus 
\mathcal{I}_{\mathcal{H}_2}$ where $\oplus$ denotes the direct
sum and $\mathcal{I}_{\mathcal{H}_j}$ denotes the identity on $\mathcal{H}_j$.

Many methods for characterizing noise channels $\mathcal{E}_{g,t}$ assume
$\mathcal{E}_{g,t}$ is trace-preserving. However, an important limitation of
many experimental implementations is that errors are not trace-preserving,
that is, generally the probability $\tr[\mathcal{P}_{\mathcal{H}_{1}}\rho]$ of 
the system being in a Hilbert space $\mathcal{H}_1\subseteq \mathcal{H}$ can 
decrease upon applying an operation, where $\mathcal{P}_{\mathcal{H}_{1}} = 
\mathcal{I}_{\mathcal{H}_{1}} \oplus 0$ is the projector onto 
$\mathcal{H}_1$~\cite{Barends2014}. 

We define the survival rate of a state $\rho\in\mathcal{H}_1$ under a CP map 
$\mathcal{E}$ to be
\begin{align}\label{def:survival_rate}
s(\rho|\mathcal{E},\mathcal{H}_1) = 
\frac{\tr[\mathcal{P}_{\mathcal{H}_1}\mathcal{E}(\rho)]}{\tr\rho}.
\end{align}

We will consider survival rates averaged over states in both $\mathcal{H}$
and in a subspace $\mathcal{H}_1$ of $\mathcal{H}$. In order to define these 
averages, note that any $\rho\in\mathcal{B}(\mathcal{H}_1)$ can be written as 
$p\tau$ for some $p\in[0,1]$ and $\tau\in\mathcal{B}(\mathcal{H})$ such that 
$\tr \tau=1$. Substituting this into Eq.~\eqref{def:survival_rate} gives
\begin{align}
s(\rho|\mathcal{E},\mathcal{H}_1) = 
\tr[\mathcal{P}_{\mathcal{H}_1}\mathcal{E}(\tau)],
\end{align}
which is a linear function of $\tau$. Consequently, the average survival rate 
in $\mathcal{H}_1$ over any measure ${\rm d}\tau$ over mixed states that is 
invariant under unitaries acting on $\mathcal{H}_1$ is
\begin{align}
s(\mathcal{E},\mathcal{H}_1) &= \int {\rm d}\tau 
\tr[\mathcal{P}_{\mathcal{H}_1}\mathcal{E}(\tau)] \notag\\
&=\tr[\mathcal{P}_{\mathcal{H}_1}\mathcal{E}(d_{\mathcal{H}_1}^{-1}\mathcal{P}_{\mathcal{H}_1})],
\end{align}
where we have used the fact that $\int {\rm d}U U\tau 
U\ct=d_{\mathcal{H}_1}^{-1}\mathcal{P}_{\mathcal{H}_1}$ for any density 
operator $\tau$, where ${\rm d}U$ is the Haar measure over unitaries acting on 
$\mathcal{H}_1$ and $d_1$ is the dimension of $\mathcal{H}_1$.

Since CP maps are linear and all quantum states can be written as $p\rho$ for 
some $p\in[0,1]$ and $\rho\in\mathcal{B}(\mathcal{H})$ such that $\tr \rho=1$, 
the survival rate for $\mathcal{H}_1=\mathcal{H}$ is strictly nonincreasing 
under composition, that is, 
$s(\rho|\mathcal{E}'\circ\mathcal{E},\mathcal{H}_1)\leq 
s(\rho|\mathcal{E},\mathcal{H}_1)$ for all CP maps $\mathcal{E}'$. In contrast, 
if $\mathcal{H}_1\subsetneq \mathcal{H}$, the survival rate can increase if 
$\mathcal{E}$ has coherences between $\mathcal{H}_1$ and $\mathcal{H}_2$. We 
therefore define the coherent and incoherent survival rates to be
\begin{align}
s_{\rm coh.}(\mathcal{E}) &= 
\tr[\mathcal{I}_{\mathcal{H}_1}\mathcal{E}(d_{\mathcal{H}_1}^{-1}
\mathcal{I}_{\mathcal{H}_1})] +
\tr[\mathcal{I}_{\mathcal{H}_2}
\mathcal{E}(d_{\mathcal{H}_2}^{-1}\mathcal{I}_{\mathcal{H}_2})]	\notag\\
s_{\rm inc.}(\mathcal{E}) &= 
\tr[\mathcal{E}(d_{\mathcal{H}}^{-1}\mathcal{I}_{\mathcal{H}})]
\end{align}
respectively. We will generally omit the argument as it will be clear from the 
context. Incoherent and coherent leakage rates can then be defined as 
$l_{\rm inc.}(\mathcal{E}) = 1-s_{\rm inc.}(\mathcal{E})$ and $l_{\rm 
coh.}(\mathcal{E}) = s_{\rm inc.}(\mathcal{E}) - s_{\rm 
coh.}(\mathcal{E})$ respectively.

\textit{Experimental protocol}---We now present a protocol for characterizing 
the average survival rates 
\begin{align}
\lvert \mathcal{G} 
\rvert^{-1}\sum_{g\in\mathcal{G}} s_{\rm inc.}(\mathcal{E}_g) &= s_{\rm 
inc.}(\mathcal{E}) \notag\\
\lvert \mathcal{G} 
\rvert^{-1}\sum_{g\in\mathcal{G}} s_{\rm coh.}(\mathcal{E}_g)
&= s_{\rm coh.}(\mathcal{E})
\end{align}
over a set of operations 
$\mathcal{G}=\{g=v\oplus (\pm w):v\in\mathcal{V},w\in\mathcal{W}\}$, where 
$\mathcal{V}$ and $\mathcal{W}$ are unitary 1-designs~\cite{Dankert2009} on 
$\mathcal{H}_1$ and $\mathcal{H}_2$ respectively, $\mathcal{E} = \lvert \mathcal{G} 
\rvert^{-1}\sum_{g\in\mathcal{G}} \mathcal{E}_g$ and the equalities follow 
from the linearity of the survival rates. Note that for incoherent leakage, 
$\mathcal{G}$ is simply a unitary 1-design on $\mathcal{H}$ such as, for 
example, the Paulis. (Note that standard RB requires a unitary 2-design, which 
is a strictly stronger requirement.) To account for weak gate-dependencies, we 
define
\begin{align}
	\Delta &= \av{\mathcal{G}}^{-1}\sum_{g\in\mathcal{G}} 
	g\circ\mathcal{E}_{g} - 
	\bar{\mathcal{G}}\circ\mathcal{E}.
\end{align}
where $\bar{\mathcal{G}} = 
\av{\mathcal{G}}^{-1}\sum_{g\in\mathcal{G}}g$, and observe that the
average variation of errors over the gate set are bounded by
\begin{align}
\epsilon &= \norm{\Delta}_{\diamond}.
\end{align}
Also note that for brevity, we assume that the noise is 
time-independent, though results for time-dependent noise can be obtained by 
applying the approaches of Ref.~\cite{Wallman2014}. 

\begin{itemize}
	\item[1.] Choose a random sequence ${\bf k} = (k_1,\ldots, 
	k_m)\in\mathbb{N}_{a}^m$ of $m$ integers uniformly at random, where 
	$\mathbb{N}_{a} = \{1,\ldots,a\}$ and $a = \av{\mathcal{G}}$.
	\item[2.] Estimate the probability $p_{\bf k}$ of detecting the system in 
	the subspace $\mathcal{H}_1$ (i.e., measuring 
	$\mathcal{I}_{\mathcal{H}_1}$) after preparing the state $\vert 0\rangle$ 
	and applying the sequence $g_{k_m}g_{k_{m-1}}\ldots g_{k_1}$ of gates.
	\end{itemize}
(Note that in standard RB, an inverse gate is applied immediately prior to the 
measurement.)

Averaging the results over a number of random sequences with fixed $m$ will 
give an estimate of
\begin{align}
\mathds{E}_{\bf k}p_{\bf k} = As_{\rm inc.}^{m-1}(\mathcal{E}) + O(m\epsilon)
\end{align}
for $\mathcal{H}_1=\mathcal{H}$ or
\begin{align}\label{eq:coherent_fit}
\mathds{E}_{\bf k}p_{\bf k} = B\lambda_{+}^{m-1} + C\lambda_{-}^{m-1} + O(m\epsilon)
\end{align}
for $\mathcal{H}_1\subsetneq\mathcal{H}$, where the constants $A$, $B$ and $C$ 
are determined by state-preparation and measurement errors (SPAM) and the 
$\lambda_{\pm}$ are fit parameters that give the coherent survival probability 
through $s_{\rm coh.}(\mathcal{E}) = \lambda_{+} + \lambda_{-}$. If the noise 
is trace-preserving on $\mathcal{H}$ (that is, if $s_{\rm inc}(\mathcal{E}) = 
1$), then Eq.~\eqref{eq:coherent_fit} simplifies to
\begin{align}
\mathds{E}_{\bf k}p_{\bf k} = B(s_{\rm coh.}(\mathcal{E})-1)^{m-1} + C + 
O(m\epsilon).
\end{align}
The gate dependent terms in these expressions are negligible provided 
$m\epsilon \ll 1$. Fitting the relevant decay curves then gives an estimate of 
the survival rates.

\textit{Derivation of the fit models.}---For the remainder of this paper we 
will work exclusively in the Liouville (or superoperator) representation of 
quantum channels, which we now briefly review. The Liouville representation is 
defined relative to a trace-orthonormal operator basis $\mathcal{A} = 
\{A_1,\ldots,A_{d^2}\}$ for the operator space $\mathcal{H}_{d^2}$ (i.e., $\tr 
A_i\ct A_j=\delta_{i,j}$). Density operators $\rho$ and measurement outcomes 
$M$ are represented by column and row vectors $|\rho)$ and $(M|$ whose $i$th 
elements are $\tr(A_i\ct \rho)$ and $\tr(M\ct A_i)$ respectively. The Born rule 
can then be expressed as $\tr M\rho = (M|\rho)$. Quantum channels (that is, 
completely positive maps) $\mathcal{E}:\mathcal{H}_{d^2}\to\mathcal{H}_{d^2}$ 
are represented by matrices $\boldsymbol{\mathcal{E}}$ such that 
\begin{align}
	\boldsymbol{\mathcal{E}}_{i,j} = \tr [A_i\ct \mathcal{E}(A_j)],
\end{align}
where $\boldsymbol{\mathcal{E}}|\rho) = 
|\mathcal{E}[\rho])$ for 
all $\rho$.

The primary advantage of using the Liouville representation is that channels 
compose via matrix multiplication, so that the probability for a sequence ${\bf k}$ is
\begin{align}
	p_{\bf k} &= (E|\boldsymbol{g}_{k_m}\boldsymbol{\mathcal{E}}_{g_{k_m}}\ldots 
	\boldsymbol{g}_{k_1}\boldsymbol{\mathcal{E}}_{g_{k_1}}|\rho),
\end{align}
where $E$ and $\rho$ are the experimental POVM elements and density matrices 
respectively. The average probability over all sequences of length $m$ is
\begin{align}\label{eq:average_operator}
	\mathds{E}_{\bf k}p_{\bf k} &= 
	\av{\mathcal{G}}^{-m}\sum_{k\in\mathbb{N}_{\av{\mathcal{G}}}^m} 
	(E|\boldsymbol{g}_{k_m}\boldsymbol{\mathcal{E}}_{g_{k_m}}\ldots 
	\boldsymbol{g}_{k_1}\boldsymbol{\mathcal{E}}_{g_{k_1}}|\rho)  \notag\\
	&= (E|\asq{\bar{\boldsymbol{\mathcal{G}}}\boldsymbol{\mathcal{E}} + 
	\boldsymbol{\Delta}}^m|\rho)	\,.
\end{align}

Since $\mathcal{G}$ is a group,
\begin{align}
	\bar{\boldsymbol{\mathcal{G}}}^2 = 
	\av{\mathcal{G}}^{-2}\sum_{g,h\in\mathcal{G}}\boldsymbol{g}\boldsymbol{h} = 
	\av{\mathcal{G}}^{-2}\sum_{g',h'\in\mathcal{G}}\boldsymbol{g}' = 
	\bar{\boldsymbol{\mathcal{G}}},
\end{align}
so the average probability simplifies to
\begin{align}
\mathds{E}_{\bf k}p_{\bf k} &= 
(E|\asq{\bar{\boldsymbol{\mathcal{G}}}\boldsymbol{\mathcal{E}}\bar{\boldsymbol{\mathcal{G}}}}^{m-1}|\rho')
 + 
\delta	\,,
\end{align}
where $\rho' = \mathcal{E}(\rho)$ and $\delta$ is the sum of all terms with 
nonzero powers of $\Delta$ obtained by expanding 
Eq.~\eqref{eq:average_operator}. Given $\epsilon$ as defined above, $\delta = 
O(m\epsilon)$, which will be negligible in practice provided $m\epsilon\ll 1$.

In order to complete the derivations, we now appeal to special properties of 
the groups $\mathcal{G}$ chosen to characterize incoherent and coherent leakage 
rates respectively.

To characterize incoherent leakage, $\mathcal{G}$ is chosen to be a unitary 
1-design, so that (see, e.g., Proposition~1 of Ref.~\cite{Wallman2014}),
\begin{align}
\bar{\boldsymbol{\mathcal{G}}} = \av{\mathcal{G}}^{-1}\sum_{g\in\mathcal{G}}\boldsymbol{g} = 
|A_1)(A_1|,
\end{align}
where $A_1 = d^{-1/2}\unit_d$, which is obtained by noting that 
the only operators invariant under conjugation by a unitary 1-design (which 
correspond to a 1-dimensional irreducible representation) are scalar matrices. 
Therefore 
\begin{align}
\bar{\boldsymbol{\mathcal{G}}}\boldsymbol{\mathcal{E}}\bar{\boldsymbol{\mathcal{G}}}
 &= |A_1)(A_1|\boldsymbol{\mathcal{E}}|A_1)(A_1| \notag\\
&= \tr\asq{A_1\mathcal{E}(A_1)}|A_1)(A_1|\notag\\
&= s_{\rm inc.}(\mathcal{E})|A_1)(A_1|
\end{align}
and so the expectation over random sequences is
\begin{align}
\mathds{E}_{\bf k}p_{\bf k} &= As_{\rm inc.}^{m-1}(\mathcal{E})	+ O(m\epsilon)	
\,, 
\end{align}
as claimed, where 
\begin{align}
 A = (E|A_1)(A_1|\rho') = \frac{1}{d}\tr E\,\tr\asq{\mathcal{E}(\rho)} \,.
\end{align}

To characterize coherent leakage, $\mathcal{G}$ is chosen so that any element 
$g\in\mathcal{G}$ can be written as $g=v\oplus \mu w$ for 
$\mu\in\{+,-\}$, where $v$ and $w$ are elements of unitary 1-designs 
$\mathcal{V}$ and $\mathcal{W}$ on $\mathcal{H}_1$ and $\mathcal{H}_2$ 
respectively. Then, using the matrix basis $\vert i\rangle\langle j\vert$ for 
the operator space, so that $\boldsymbol{U} = U\otimes U^*$ where $^*$ denotes 
complex conjugation, we have (again by Proposition~1 of Ref.~\cite{Wallman2014})
\begin{align}
\bar{\boldsymbol{\mathcal{G}}} &= \av{\mathcal{G}}^{-1}\sum_{\mu,v,w} 
(v\oplus \mu w) \otimes (v\oplus \mu w)^*\notag\\
&= \big(\av{\mathcal{V}}^{-1}\sum_{v} v\otimes v^*\big) \oplus 
0\oplus 0 \oplus \big(\av{\mathcal{W}}^{-1}\sum_w w\otimes w^*\big)
\notag\\
&=|d_1^{-1/2}\mathcal{P}_{\mathcal{H}_1})(d_1^{-1/2}\mathcal{P}_{\mathcal{H}_1}|
 + 
|d_2^{-1/2}\mathcal{P}_{\mathcal{H}_1})(d_2^{-1/2}\mathcal{P}_{\mathcal{H}_1}|.
\end{align}

Setting $A_1 = d_1^{-1/2}\mathcal{P}_{\mathcal{H}_1}$ and $A_2 = 
d_2^{-1/2}\mathcal{P}_{\mathcal{H}_2}$, we then have
\begin{align}
\bar{\boldsymbol{\mathcal{G}}}\boldsymbol{\mathcal{E}}\boldsymbol{\bar{\mathcal{G}}} = s\oplus 0,
\end{align}
where $s$ is a $2\times 2$ matrix. We can easily take powers of $s$ by putting 
it in lower-triangular form, so that
\begin{align}
\mathds{E}_{\bf k}p_{\bf k} &=	B\lambda_+^{m-1} + C\lambda_-^{m-1}	+ O(m\epsilon)
\end{align}
where
\begin{align}
\lambda_{\pm}=\frac{s_{1,1} + s_{2,2}}{2} \pm 
\frac{1}{2}\sqrt{(s_{1,1}-s_{2,2})^2 + 
4s_{1,2}s_{2,1}}.
\end{align}
are the eigenvalues of $s$ and $B$ and $C$ are constants (which absorb both the 
SPAM and the unitary that makes $s$ lower-triangular).

The sum of the eigenvalues is equal to 
$s_{1,1} + s_{2,2}=s_{\rm coh.}(\mathcal{E})$ since
\begin{align}
s_{1,1} &= (A_1|\boldsymbol{\mathcal{E}}|A_1) = 
\tr\asq{\mathcal{P}_{\mathcal{H}_1}\mathcal{E}(\frac{1}{d_1}\mathcal{P}_{\mathcal{H}_1})}
	\notag\\
s_{2,2} &= 
\tr\asq{\mathcal{P}_{\mathcal{H}_2}\mathcal{E}(\frac{1}{d_2}\mathcal{P}_{\mathcal{H}_2})}.
\end{align}
If the noise is trace-preserving, then one of the eigenvalues must be one 
(corresponding to $\unit_{d+e}$), and the other must then be 
$s_{\rm 
coh.}(\mathcal{E}) - 1$.

\begin{figure}[t]
	\centering
	\includegraphics[width=\linewidth,keepaspectratio]{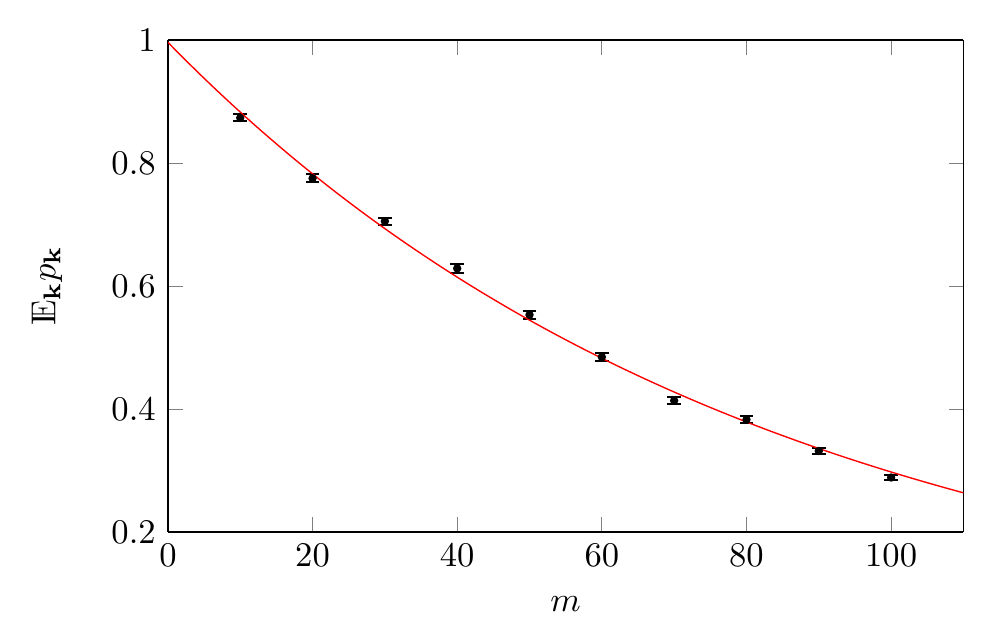}
	\caption{(Color online) Numerical fidelity decay curve for the 
		gate-dependent noise on single-qubit Paulis in 
		Eq.~\eqref{eq:incoherent_channel}. The data points are the estimates of 
		$\mathds{E}_{\bf k}p_{\bf k}$ for $m = 10, 20, ..., 100$ obtained by 
		sampling 30 random sequences of single-qubit Pauli operators and the 
		error bars are the standard errors of the mean. The colored line is the 
		fit to the model obtained using MATLAB's nlinfit package, which gave 
		$s_{\rm inc.}(\mathcal{E}_{\rm fil}) = 0.9880(2)$
		with an $r^{2}$ value of 0.9991, compared to the theoretical value of 
		$s_{\rm inc.}(\mathcal{E}_{\rm fil}) = 0.9879$, where $\mathcal{E}_{\rm 
		fil} = \sum_i \mathcal{E}_i/4$ is the average weakly filtering channel.}
	\label{fig:incoherent}
\end{figure}

\begin{figure}[t]
	\centering
	\includegraphics[width=\linewidth,keepaspectratio]{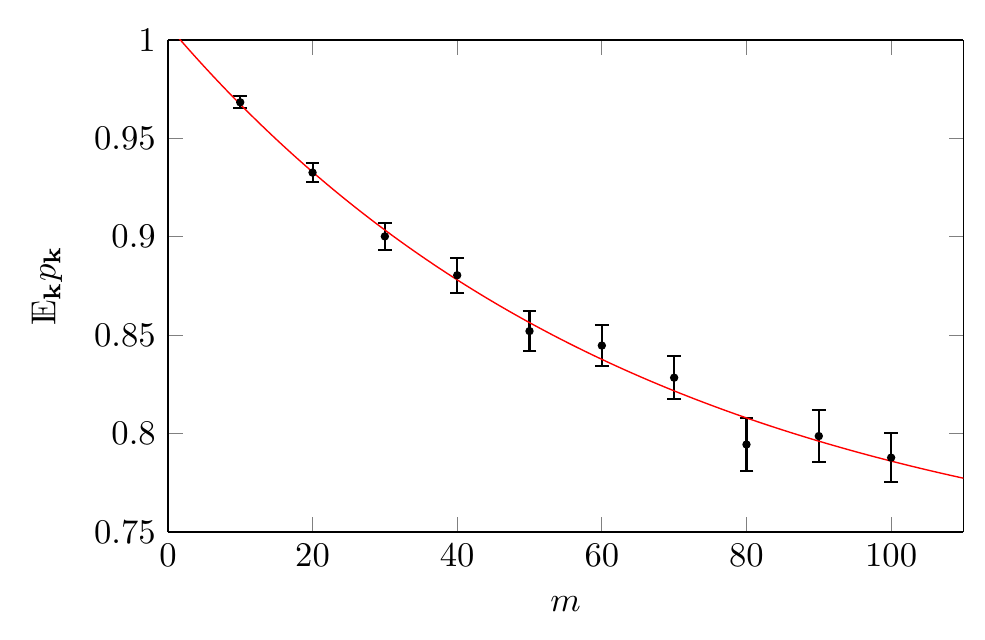}
	\caption{(Color online) Numerical fidelity decay curve for the composite 
		noise channel in Eq.~\eqref{eq:coherent_noise}. The data points are the 
		estimates of $\mathds{E}_{\bf k}p_{\bf k}$ for $m = 10, 20,..., 100$ 
		obtained by sampling 200 random sequences of qutrit operators in 
		$\mathcal{G}$ and the error bars are the standard errors of the mean. 
		The colored line represents the fit to the model obtained using 
		MATLAB's nlinfit package, which gave $s_{\rm 
		coh.}(\overline{\mathcal{E}}_X) = 0.992(2)$ with an $r^{2}$ value of  
		0.9904, compared to the theoretical value 
		$s_{\rm coh.}(\overline{\mathcal{E}}_X)= 0.995$, where 
		$\overline{\mathcal{E}}_X$ is the channel obtained by integrating 
		$\mathcal{E}_X$ over the appropriate measures.}
	\label{fig:shelving}
\end{figure}

\textit{Numerical simulations.}---Results of numerical simulations of our 
protocol for two specific models of incoherent and coherent leakage are 
illustrated in figures~\ref{fig:incoherent} and \ref{fig:shelving} respectively,
demonstrating robust performance with a model of (weakly) gate-dependent 
errors.  

For the numerical simulations of our protocol for incoherent leakage, the set 
of operations $\mathcal{G}$ is the set of single-qubit Paulis and we modeled 
the gate-dependent error $\mathcal{E}_i$ for each $g_i$ as
\begin{align}\label{eq:incoherent_channel}
\mathcal{E}_i(\rho) = \frac{p_i}{4}(\unit + r_i\cdot\vec{\sigma})\rho(\unit + 
r_i\cdot\vec{\sigma}) + (1-p_i)\rho,
\end{align}
where $\vec{\sigma}=(X,Y,Z)$ is the vector of Paulis and $p_i\in[0,0.05]$ and  
$r_i\in\mathcal{S}^2$ (the unit sphere) were chosen independently and uniformly 
from the appropriate measures. The channels $\mathcal{E}_i$ correspond to
channels that weakly filter out (that is, absorb) the component of a state 
orthogonal to some randomly-chosen state with Bloch vector $r_i$. When $r_i$ is
fixed, this corresponds to loss from a particular energy level. However, we 
chose $r_i$ randomly to accentuate the statistical fluctuations as much as 
possible.

For numerical simulations of our protocol for coherent leakage, we adopted  a 
noise model that is motivated by experimental techniques that use an auxiliary 
level (e.g., ``shelving'' in ion trap experiments~\cite{Blatt2004}) to protect 
certain states while performing another operation. The ideal shelving gate is a 
Pauli $X$ rotation between the second and third level, that is, $V_{ideal} 
= 1 \oplus X$. The group $\mathcal{G}$ of operations is $\{P\oplus \pm 
1:P=\unit,X,Y,Z\}$. Our model of coherent leakage at each time step is 
\begin{align}\label{eq:coherent_noise}
\mathcal{E}_X = V_{\gamma_2}\circ \delta U_2 \circ V_{\gamma_1} \circ 
\delta U_1,
\end{align}
where
\begin{align}\label{eq:shelving_operator}
V_{\gamma} &= 1 \oplus \left(\begin{array}{cc}
i\sin\gamma & \cos\gamma\\
\cos\gamma & i\sin\gamma
\end{array} \right)	\notag\\
\delta U &= e^{i\phi U X U\ct}\oplus 1.
\end{align}
That is, our noise model consists of imperfect shelving ($V_{\gamma_1}$) and 
unshelving ($V_{\gamma_2}$) gates, together with some small coherent noise on 
the code space ($\delta U_1$ and $\delta U_2$). The channel $\mathcal{E}_X$ is 
trace-preserving on the combined code and leakage space, but is 
trace-decreasing when restricted to the code space. 
We sampled $U$ from the Haar measure on the code space with $\phi = 0.01$ 
fixed and $\gamma$ from the normal distribution with zero mean and $\sigma = 
0.06$, where each variable is sampled independently each time the relevant gate 
is applied.

\textit{Conclusion}---In this paper, we have presented a protocol for 
characterizing average survival rates under incoherent leakage and coherent 
leakage to an orthogonal subspace. Experimentally implementing our protocol 
yields a decay curve which can be fitted to our analytical expressions 
to obtain the average probability of a leakage event occurring. If the
experimental data deviates significantly from our decay curves, then 
the experimental noise is either strongly gate-dependent or non-Markovian.
We have also demonstrated that the decay can be observed and fitted in practice 
through numerical simulations of leakage for specific error models.

Our protocol is scalable and robust against state-preparation and measurement 
errors. Our current protocol can also be applied in conjunction with standard 
RB to determine both the average leakage rate and the average gate infidelity 
over a unitary 2-design such as the Clifford group.

As with standard RB, obtaining rigorous confidence intervals on the parameters 
obtained from our protocol is still an open problem, though techniques bounding 
the number of sequences to be sampled~\cite{Wallman2014} and using Bayesian 
methods to refine prior information~\cite{Granade2014} should also be 
applicable to our protocol.

\textit{Acknowledgments}---The authors acknowledge helpful discussions with 
S.~Flammia, C.~Granade and T.~Monz. This research was supported by the U.S. 
Army Research Office through grant W911NF-14-1-0103, CIFAR, the Government of 
Ontario, and the Government of Canada through NSERC and Industry Canada.

\end{document}